# IMPLEMENTATION OF AN AUTOMATIC SYLLABIC DIVISION ALGORITHM FROM SPEECH FILES IN PORTUGUESE LANGUAGE

E.L.F. DA SILVA AND H.M. DE OLIVEIRA

*Federal University of Pernambuco, UFPE, Brazil*

lizandra_fernandes@hotmail.com, hmo@ufpe.br

**Abstract**— A new algorithm for voice automatic syllabic splitting in the Portuguese language is proposed, which is based on the envelope of the speech signal of the input audio file. A computational implementation in Matlab™ is presented and made available at the URL http://www2.ee.ufpe.br/codec/divisao_silabica.html. Due to its straightforwardness, the proposed method is very attractive for embedded systems (e.g. i-phones). It can also be used as a screen to assist more sophisticated methods. Voice excerpts containing more than one syllable and identified by the same envelope are named as super-syllables and they are subsequently separated. The results indicate which samples corresponds to the beginning and end of each detected syllable. Preliminary tests were performed to fifty words at an identification rate circa 70% (further improvements may be incorporated to treat particular phonemes). This algorithm is also useful in voice command systems, as a tool in the teaching of Portuguese language or even for patients with speech pathology.

## 1. Introduction

With the advent of more efficient circuitry and the increasing availability of new signal processing techniques (Vasegnhi 2007, Holmes & Holmes, 2001, Oppenheim & Schafer, 2010), the development of new apps involving speech synthesis, speaker recognition, automatic translators for speech conversion (text-to-speech and / or spoken text), are proliferating (Taylor, 2003, Santos, 1997 Sotero & de Oliveira, 2009). Applications systems involving of speech-to-text conversion in Portuguese are already promising (Fraga, 2001, Silva et al., 2008, Silva et al., 2009), but the current systems still require improvements (Neto, 2005). Some systems intended for Portuguese text-to-speech have been suggested (e.g. (Fraga, 2001)) and they typically involve three steps: *i*) (sub)syllabic segmentation; *ii*) conversion of segmented phonemes into text; and *iii*) spellchecking and grammar of words and sentences identified (Huang et al. 2001). The emphasis of this work focus on the step of segmenting speech signals with unlimited vocabulary for a automatic hyphenation (Gouveia et al., 2000). Rosenberg and colleagues (Rosenberg et al. 1983) proposed a system for word recognition by concatenating half-syllables defined previously, using reference models (templates) with error rates in the recognition of subsyllables in the range 18-33%. Other systems for isolated words recognition were proposed, which are based on the "dynamic time warping" (Lipeika et al. 2002). Often, the segmentation employing automatic speech models in which phonetic subunits (context dependent) are represented using Hidden Markov Models-HMM (Selmini, 2008, dos Santos & Alcaim, 2001). Although the success rates of such systems are acceptable, the algorithms are computationally intensive, especially for real time applications or in the development of embedded systems (iPods, iPhones, etc.). The alternative technique offered here is quite simple (compared to the HMM techniques or Cepstral analysis involved), which may work as a preliminary step to be incorporated into existing algorithms. This is not a competition or a direct alternative to existing systems (neither hit rates nor complexities are at the same level), but in a way to achieve some pre-processing. Then, after the initial application of this technique, further strategies may be incorporated. Despite the modest goal, this method can increase the speed of speech-to-text (for Portuguese) already proposed. After this brief introduction that laying the subject, Section 2 describes the acquisition and pre-processing speech. In particular, it is described implementing a procedure for the identification the envelope of the speech signal based on a linear discharge envelope detector. Section 3 presents a proposal for location of syllables in the audio signal. The identification of supersyllables and his "breaking" in smaller syllables are described in Section 4. Finally, some validation tests of the proposed system are presented along with a preliminary analysis of the performance the syllabic splitter, besides the conclusions.

## 2. Acquisition and pre-processing

### 2.1. Acquisition and Pre-processing Speech

Data (speech) were collected in an isolated room with controlled noise level. The compilation was performed with the aid of a Intel HP® Atom computer, unused headset. There were performed 49 distinct speech collections for males and female. Software used to acquire the speech was Audacity 1.3™. The recordings generated files with ".wav" extension. Figure 1 illustrates the interface of the Matlab script for the file audio corresponding to the word "*departamento*". The preprocessing the vocalizations was also carried out with the aid of the same application, which consists in eliminating the quiet portions at the beginning and end of each recorded file.

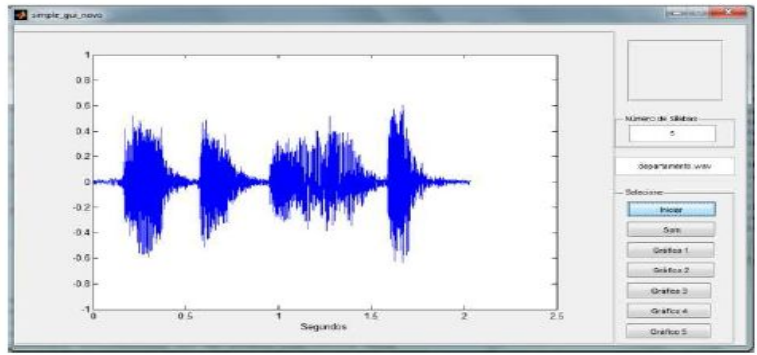

*Figure 1: Illustration of the Matlab program interface for word "departmento" after preprocessing.*

Because the files could be acquired stereo format, it may be required a conversion on the Mono format. The reading syllabic separator is then performed at a rate sampling of 44.1 kHz, with a 16 bits quantization. The window length in the algorithm was 2,048 samples. This value was selected by examining the vocal behaviour of speakers of Portuguese (Fernandes da Silva & de Oliveira, 2012).

### 2.2 Rectification of the wave, RMS value and Linear Discharge

In this step, first, it takes place a half-wave rectification of the input signal. The rectification the speech signal is performed with the aid of the function signal, defined by:

$$sign(x) = \begin{cases} -1 & se\ x < 0 \\ 0 & se\ x = 0 \\ 1 & se\ x > 0 \end{cases} \quad (1)$$

wherein each value of sample $x[n]$ is converted for $x[n].(1+sign(x[n]))/2$. Figure 2 illustrates the GUI Matlab program for signal wave rectified for the word "batata" (adopted in illustration instead of the word "*departmento*" for the sake of simplicity). After performing the half-wave rectification, the rms (root mean square) was assessed.

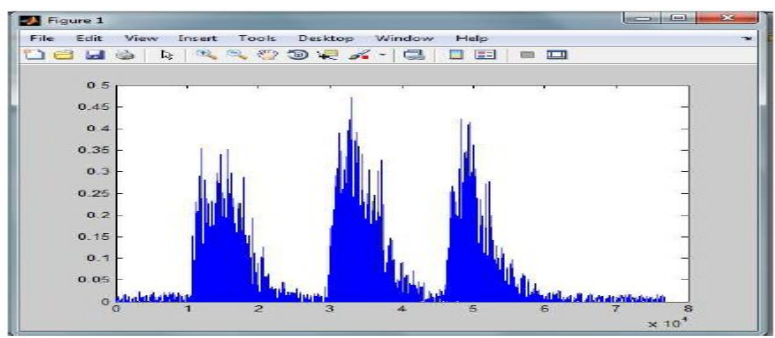

*Figure 2: Illustration of the Matlab GUI for the signal audio of the word "batata" after half-wave rectification.*

The rms value is used later in the process of the syllabic location. Along with a statistical percentage, both serve as a criterion for assessment on the location of the syllables. The envelope detection is then performed through a linear discharge of the signal. For this, values of samples obtained in the wave-rectification process are used. This process consists of describe the envelope of the rectified wave through a linear predictive method (Juang & Rabiner, 1993). The samples are analysed within a subwindow linear discharge (referred to as delta) within a larger window, which has 2,048 samples. Delta is a value obtained by the product of the sampling rate (44.1 kHz) by the value of the vocoders time-window, which is typically 22 ms (Rabiner & Schafer, 2007). The result of this product provides a small window with 970 samples that describes the trajectory of the discharge (approximate linearly) peak, aiming to pursue the signal envelope. The delta value corresponds to the adjusting of the time constant of the AM detector (de Oliveira, 2012). Illustration of the signal wave rectified can be seen in Figure 3.

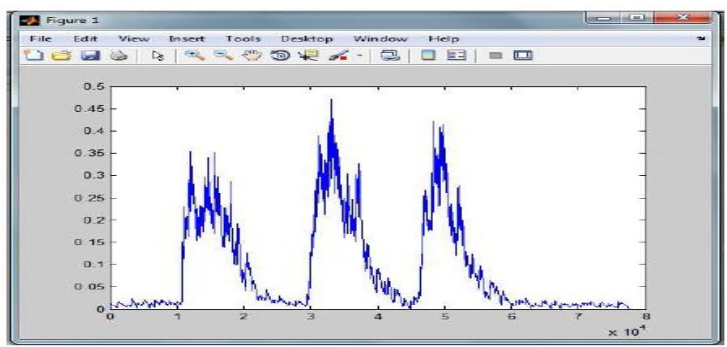

*Figure 3: Illustration of the graphical Matlab interface to the envelope audio signal of the word "batata".*

## 3. The syllable finder

The location of the syllables in the audio signal is carried out by comparing the value of each sample (obtained from the envelope detection step) with a threshold value. This threshold was set after several empirical tests conducted by successive analyses. The threshold was expressed in percentage (set value was Perc = 1.2) in terms of the rms value obtained for half-wave rectification step. The algorithm consists of assigning the value "1", a new vector (named "syllable-vector") in its position in the sample value. This assignment occurs when the sample value of the envelope is greater than a selected threshold value. Otherwise, it is assigned the value "0". Then, counting of "zeros and ones" in the stream of the syllable-vector.

The count of consecutive sampling sequences is performed and then the values obtained are stored in the array named "matrix metering". This matrix stores the number elements found in the sequence (second column) and its associated value (first column). The count array can be seen in Table I for the separation of the word "batata" as illustration (N.B. this is merely explanatory, one since a better notation can be adopted as will be seen). The next step is to remove the very short sample sequences mapped to the value 1, in order that such short sequences do not interfere with the location of syllables within every word. Samples of short sequences are eliminated by comparison with a percentage. This percentage was also obtained by *ad hoc* adjustments and corresponds to 1.8 times the length of the window used to describe the trajectory of the envelope (970 samples).

*Table I: Matrix metering obtained via count Matlab to file extract audio from Pre-processed concerning the word "batata".*

| value in sequence | amount elements |
|---|---|
| 0 | 10702 |
| 1 | 9472 |
| 0 | 204 |
| 1 | 788 |
| 0 | 185 |
| 1 | 40 |
| 0 | 55 |
| 1 | 92 |
| 0 | 8238 |
| 1 | 29 |
| 0 | 192 |
| 1 | 9432 |
| 0 | 523 |
| 1 | 484 |
| 0 | 128 |
| 1 | 10 |
| 0 | 423 |
| 1 | 46 |
| 0 | 512 |
| 1 | 156 |
| 0 | 4633 |
| 1 | 8631 |
| 0 | 189 |
| 1 | 611 |
| 0 | 1137 |
| 1 | 79 |
| 0 | 68 |
| 1 | 24 |
| 0 | 19717 |

A more compact representation may be used, simplifying the data to a $0_n 1_m$ notation, or where the indices denote the lengths of sequences of "zeros" and "ones" respectively. The count array displayed in Table I can be represented by the sequence (runlength) of alternating 0's and 1's:

$$0_{10702} 1_{9472} 0_{204} 1_{788} 0_{185} 1_{40} 0_{55} 1_{92} 0_{8238} 1_{29} 0_{192} 1_{9432} 0_{523} 1_{484} 0_{128} 1_{10} 0_{423} 1_{46} 0_{512} 1_{156} 0_{4633} 1_{8631} 0_{189} 1_{611} 0_{1137} 1_{79} 0_{68} 1_{24} 0_{19717}$$

After the purging of the short sequences of elements "1" (false syllable location), it yields on:

$$0_{10702} 1_{9472} 0_{204} 0_{788} 0_{185} 0_{40} 0_{55} 0_{92} 0_{8238} 0_{29} 0_{192} 1_{9432} 0_{523} 0_{484} 0_{128} \, 0_{10} 0_{423} 0_{46} 0_{512} 0_{156} 0_{4633} 1_{8631} 0_{189} 0_{611} 0_{1137} 0_{79} 0_{68} 0_{24} 0_{19717}$$

i.e. $0_{10702} 1_{9472} 0_{9823} 1_{9432} 0_{6915} 1_{8631} 0_{21825}$.

At the end of this step, we make the summation the number of positions that have only values unit in order to obtain the total number of syllables. In the current example, three syllables were identified with well delineated start-and-end.

As seen, the elements in the counting sequence establish which samples contain the syllables of the analysed word. These same unit values are also responsible for determining the beginning of each syllable. To find out the location of each one of them a procedure is done in which, first, it is computed the *cumulative sum* of count sequence, storing the result in a new vector - the "positions vector." Then it can be seen in which position a change of value occurs. Syllable division is obtained for the stretch of audio (word: batata) after his separation, which corresponds to 76,800 samples in the recorded audio (i.e., ~ 1.75 s duration).

$$0_{10702} 1_{9472} 1_{29997} 2_{39429} 2_{46344} 3_{54975} 3_{76800}.$$

### 4. Identifying and breaking supersyllables

Ordinarily, the average number of samples contained in a syllable is 6,543 samples, with a standard deviation of 5,467 samples (values limited to the statistical analysis of investigated words, but representative for a general line of analysis in Portuguese - At least for the purposes of this algorithm). So long syllables whose lengths exceeds the mean by one average standard deviation (i.e. 12,010 samples) are considered to be " supersyllables". With empirical basis, it was observed that the sequences of a number of samples equal to or above this value had *de facto* more than a syllable in its composition. Furthermore, it has become necessary to use other criteria to location of the start and end of each syllable within a supersyllables.

The criterion used to break up syllables containing number greater than or equal to 12,010 samples is to compare the values of samples of this supersyllable with a pre-set value. This value is the result of the product of the threshold used in location of the syllable by an adjustment factor denoted by "epsilon". The process is rather similar to that used to detect of syllables per threshold, now at another level (inspired the standard type Matrioshka - Russian dolls).

## 5. Performance of the syllabic divisor

The proposed algorithm could perform adequately an automatic syllabic separation of 43 from 49 recorded words. Among these 43 words, 20 of them were separated completely (Table III) and 23 obtained a partial separation (Table IV). It is shown in the sequel, illustrations of the process for the separation of word DEPARTMENTO, including details of the interface graphic. For these graphs (Figures 5 to 7), corresponding sounds and matching files can be found in the URL cited in the abstract.

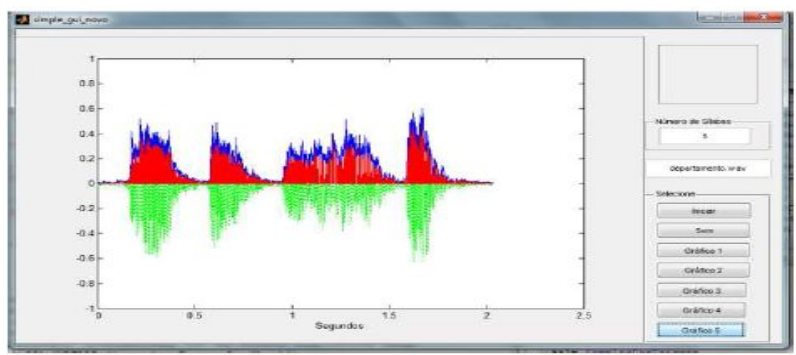

***Figure 4:*** *Waveforms involved: a) green colour signal, audio on the word "departmento"; b) in red, the half-wave rectified waveform; c) in blue, the envelope signal obtained with linear envelopment discharge.*

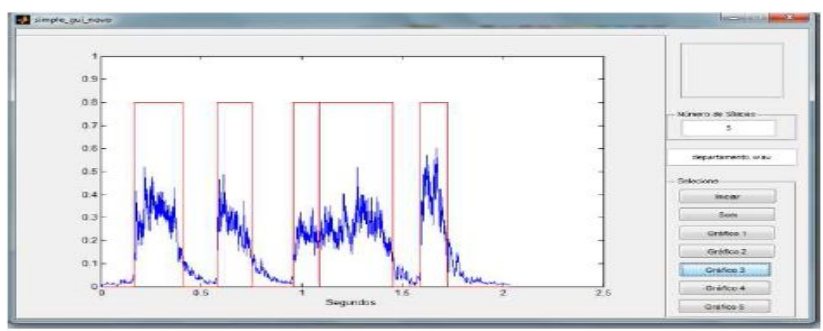

***Figure 5:*** *Waveform with its syllabic separation obtained by threshold applied to the envelope of the speech signal (pronounced word: "Departmento").*

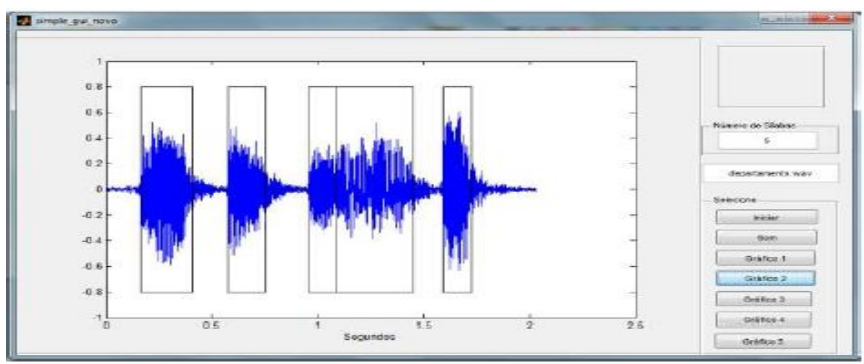

***Figure 6:*** *Separation syllable in the original signal: note the separation clear phonemes DE-PAR-TAMEN-TO. The TAMEN supersyllable was properly divided.*

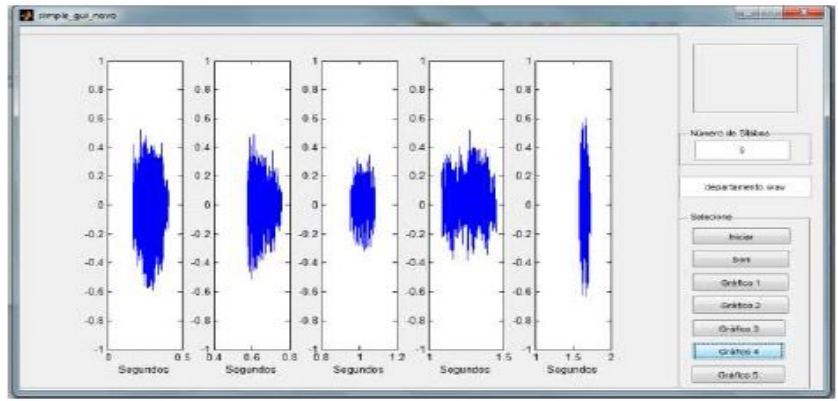

*Figure 7: Syllable Division completed. Excerpts are indicated; there access to samples in length and duration of syllables and audio snippet isolated from each syllable.*

The assessed word, "departmento", corresponded to a total of 76,800 samples in the recording, lasting total of 1.74 seconds. Table II identifies the syllabic output separator, with their proposed syllables (final and initial samples) and the estimated duration.

*Table II: Syllables in separate voice file containing the word "Departmento".*
*The index of: the initial sample, the final sample and estimated duration of the syllable (example of output data).*

| Syllable | onset | end | Duration (ms) |
|---|---|---|---|
| De | 7371 | 17913 | 239 |
| Par | 25364 | 33320 | 180 |
| Ta | 42188 | 47975 | 131 |
| Men | 47975 | 64029 | 364 |
| To | 70141 | 76111 | 135 |

*Table III: List of words with their respective (correctly) separate syllable obtained by the algorithm of syllabic division for audio archive (23 words).*

| Words | syllable |
|---|---|
| Abacate | a - ba - ca - te |
| Batata | ba - ta - ta |
| Berimboca | be - rim - bo - ca |
| Bonita | bo - ni - ta |
| Butantã | bu - tan - tã |
| Café | ca - fé |
| Campus | cam - pus |
| Complexo | com - plex - xo |
| Computador | com - pu - ta - dor |
| Corpo | cor - po |
| Departamento | de - par - ta - men - to |
| História | his - tó - ria |
| Hoje | ho - je |
| Música | mus - si - ca |
| Pitoco | pi - to - co |
| Recife | rec - ci - fe |
| Roupa | rou - pa |
| Semiconductor | se - mi -cond - du - tor |
| Siri | si - ri |
| Solteiro | sol - tei - ro |
| Uva | u - va |
| Vale | va - le |
| Zebra | ze - bra |

It may be noted that the words listed in Table IV were not entirely separate (e.g. vião, silei). Although some snippets of voice possessed more than one syllable, which did not meet the criteria used to define the supersyllables and therefore require still suffer a further decline. In other cases, single letters with a greater duration than the usual (e.g., **b**, **m**, **s**) are indicated as pseudosyllable (this can not be eliminated by changing the threshold, under penalty of "losing" a few short syllables). In a few words, it was not possible to separate syllables, for example, *oito*, *roxo* and *banana*.

Preliminary tests with several different speakers and words also indicated that an increase in speed with which the announcer pronounces the word(s) (or speech) almost does not degrade the performance of the syllable recognition scheme. An additional (rather convenient) test consisted in execution of the algorithm to an audio file obtained by reading fluent, continuous speech, a written text. The text selected was the whole poetry "*I'm off to Pasargadae*," the poet Manuel Bandeira burned to wav 16 bit mono, lasting 1 minute and a half, which contains a total 316 syllables.

*Table IV:* List of words syllabic partial separation (20 total words)

| Syllable | words |
|---|---|
| Abacaxi | a ba *cax* i |
| Assado | as sa *d* do |
| Avião | av *vião* o |
| Brasileiro | Bra *silei* ro |
| Cabelo | Ca *belo* |
| Cabeça | ca be *ç* ça |
| Circumferência | cir cun fe rên *c* ia |
| Duzentos | duz zen |
| Economia | e *cono* mia |
| Eletrônica | *ele* tro ni ca |
| Engenharia | Em *g genhar* ria |
| Farmácia | Far r ma cia |
| Matemática | Ma *tema* a ti ca |
| Minuto | M m inu to |
| Mistério | Mi s te rio |
| Oficina | O fic c *ina* |
| Pernambuco | Per nam *b* bu co |
| Televisão | *telev* v *visão* |
| Universidade | Uni ver si *dade* |
| Vestibular | Vês ti *bula* ar |

The syllabic identifier proposed found a total of 360 syllables (as including among them, 62 supersyllables). There are still circa 5% of stretches identified as syllables, but that constitute mere single letters. The inclusion of additional criteria to reduce these occurrences can improve the performance of the separator.

## 6. Conclusions

A major challenge in speech processing is the design of systems for the automatic speech-to-text (STT), with running in real time with acceptable quality (Huang et al. 2001). Systems text-to-speech (TTS) are highly developed. This algorithm can be useful in an initial stage of a TTS system adapted to Portuguese. The technique introduced here is much more simple when compared to the HMM techniques or cepstral analysis. In the case of use in embedded systems, despite the hit rate of syllables not be suitable for commercial applications, simple improvements can be incorporated for treat phonemes for which the envelope is not key for the identification. The proposal (with code open) may operate as a first base system and thus, better accuracy rate can be obtained.

A particular case of interest and application involves the use in automated systems with commands specific voice (previously catalogued). In this case, the hit rates can be much better. Another potential application is in the application aid in early childhood education or as a tool support the teaching of Portuguese for foreigners. Another particularly attractive scenario is the use as support for patients underwent speech therapy and medical diagnosis (Fagundes et al. 2008).

**Acknowledgements**

E.L.F. Silva thanks CNPq for financial support, and support the PPGEE-UFPE. The authors express gratitude to Paul Hugo Espirito Santo, Paulo Roberto Martins and Raimuno C. de Oliveira for his help and suggestions on the implementation of the Matlab code.